\documentclass [a4paper,12pt,fleqn]{article}
\usepackage{graphicx}
\usepackage {subfigure,epsfig}

\usepackage {amsmath} \usepackage{amssymb} \usepackage{cite} \usepackage{amsthm}
\numberwithin{equation}{section} \numberwithin{table}{section} \mathindent=0pt
\theoremstyle{plain} \newtheorem{theorem}{Theorem}

\numberwithin{theorem}{section}

\begin{document}

\title{Painleve property and the first integrals of nonlinear differential equations}
\author{N.A. Kudryashov}
\date{Department of Applied Mathematics\\
Moscow  Engineering and Physics Institute\\
(State university)\\
31 Kashirskoe Shosse,  115409\\
Moscow, Russian Federation} \maketitle

\begin{abstract}
Link between the Painleve property and the first integrals of
nonlinear ordinary differential equations in polynomial form is
discussed. The form of the first integrals of the nonlinear
differential equations is shown to determine by the values of the
Fuchs indices. Taking this idea into consideration we present the
algorithm to look for the first integrals of the nonlinear
differential equations in the polynomial form. The first integrals
of five nonlinear ordinary differential equations are found. The
general solution of one of the fourth ordinary differential
equations is given.

\end{abstract}



\section{Introduction}

Finding of the first integrals of the nonlinear ordinary
differential equations is one of the most important problem in the
theory of nonlinear differential equations. The first integrals
allow us to obtain the general solution of nonlinear differential
equation in the form of quadratures.

The Painleve property is known to be the necessary condition to
integrate nonlinear ordinary differential equation \cite{1,2,3}.
There is the link between the Painleve property and integrability
of nonlinear differential equation. Using the singular manifold
method many interesting properties were found \cite{4,5,6,7,8}. We
note the Lax pairs and the Backlund transformations for nonlinear
partial differential equation \cite{6,7} and the birational
transformations \cite{9,10} that can be found for nonlinear
ordinary differential equations if the nonlinear differential
equations have the Painleve property.

There are two aims of this paper. The first aim is to present the
link between the Fuchs indices in the expansion of solution in the
Laurent series and the first integrals of nonlinear ordinary
differential equations. The second aim is taking the Fuchs indices
into account to present the algorithm for finding of the first
integrals of nonlinear ordinary differential equations that have
the Painleve property.

Our approach is illustrated to find the first integrals of the
following nonlinear differential equations

\begin{equation}
\label{eq1}
 y_{xxxx} - 10y\,y_{xx} - 5y_x^2 + 10y^3 + \beta \left( {y_{xx} - 3y^2}
\right)\, + \,\delta \,y + \mu = 0
\end{equation}

\begin{equation}
\label{eq2}
 y_{xxxx} - 10y\,^2y_{xx} - 10y\,y_x^2 + 6y^5 + \beta \,\left( {y_{xx} -
2y^3} \right)\, + \delta \,y + \mu = 0\,\,\,\,\,\,\,\,\,\,\,\, \\
\end{equation}

\begin{equation}
\label{eq3}
 y_{xxxx} + 5y_x y_{xx} - 5y\,^2y_{xx} - 5y\,y_x^2 + y^5 - \delta =0
\end{equation}

\begin{equation}\begin{gathered}
\label{eq4} y_{{{\it xxxx}}}-4\,{\frac {y_{{x}}y_{{{\it
xxx}}}}{y}}+ {\frac {{21\,y_{{x}}}^{2}y_{{{\it xx}}}}{{2\,y}^
{2}}}-3\,{\frac {{y_ {{{\it xx}}}}^{2}}{y}}-5\,{\frac
{\delta\,y_{{{ \it xx}}}}{{y}^{2}}}-\,{\frac
{9\,{\,y_{{x}}}^{4}}{2\,{y}^{3}}}+10\,{\frac {\delta\,{y_{{x}}}^{2}}{{y}^{3}}}+\\
\\
+\nu \,{y}^{2}-2\,{\frac {{\delta}^{2}}{{y}^{3}}}+\mu=0
\end{gathered}\end{equation}

\begin{equation}\begin{gathered}
\label{eq5}
 y_{{{ xxxx}}}-2\,{\frac {y_{{{
xxx}}}y_{{x}}}{y}}-\,{\frac {3\,{y_{{{xx}}}}^{2}}
{2\,y}}+2\,{\frac {y_{ {{\it
xx}}}{y_{{x}}}^{2}}{{y}^{2}}}-5\,{y}^{2}y_{{{xx}}}-\frac52\,y{y_{{x}}}^{2}+\frac52\,{y}^{5}-\\
\\
-{\beta}\,{y }^{3}+\mu\,y=0
\end{gathered}\end{equation}

Equation \eqref{eq1} can be obtained from the generalized Korteveg
-- de Vries of the fifth order if we look for solution of this
equation in the form of the travelling wave. Equation \eqref{eq1}
was found as the special case at the description of the nonlinear
wave propagation on the water \cite{11}. This equation  can be
found at $\beta=0$ as equation for asymptotic solution of the
fourth order analogy for the first Painleve equation
\cite{12,13,14,15}. Equation \eqref{eq1} was studied at $\beta=0$
in several works \cite{16,17,18}.

Equations \eqref{eq2} and \eqref{eq3} can be found from the
modified  Korteveg -- de Vries and the Kaup -- Kupershmidt
equations of the fifth order if we look for solution in the form
of the travelling wave. Equation \eqref{eq3} was obtained by A P
Fordy and J Gibbons  in \cite{19}. All these equations are used as
equations for the description of the Henon - Heiles model
\cite{13,20,21}. Equations \eqref{eq4} and \eqref{eq5} are found
as special cases of equations that were obtained recently as the
higher order analogies of the Painleve equations
\cite{22,23,24,25}. All above mentioned equations have the
Painleve property.

The outline of this paper is as follows. In section 2 we present
the link between the Fuchs indices and the first integrals of
nonlinear differential equations in the polynomial form. Section 3
is devoted to the discussion of the algorithm to look for the
first integrals of nonlinear ordinary differential equations. In
sections 4, 5, 6 and 7 we give examples of finding of the first
integrals of the above mentioned nonlinear ordinary differential
equations. In section 8 we demonstrate the application of the
first integrals to look for the general solution of equation
\eqref{eq1}.

\section{Link between the Fuchs indices and the first integrals}

Assume we have the nonlinear ordinary differential equation in the
polynomial form that passes the Painleve test and assume we are
going to find the first integrals to look for the general solution
of this equation using quadratures. Taking results of the Painleve
analysis into account we have the Fuchs indices and necessary
amount of arbitrary constants in the expansion of the general
solution in the Laurent series. Our aim is using values of the
Fuchs indices to find the general form of the first integrals of
nonlinear ordinary differential equations. To understand the link
between values of the Fuchs indices and the first integrals of
nonlinear ordinary differential equations first of all let us
consider two simple examples.

\textit{Example 1.} Consider the equation

\begin{equation}\label{eq01}
E_1[y]=y_{zz}+6\,y^2- a y-\, {b}=0
\end{equation}

Equation \eqref{eq01} has the first integral in the form

\begin{equation}\label{eq02}
I_0[y]=y_{z}^2+4\,y^3- a y^2-2\,b\,y-C_1=0
\end{equation}

The general solution of equation \eqref{eq01} is expressed via the
Weierstrass function. The solution of equation \eqref{eq01}  has
one branch in the expansion in the Laurent series and the Fuchs
indices $(j_1,\,j_2)=(-1,\,6)$.

Equation possesses the Painleve property and there is the local
presentation of the general solution in the form

\begin{equation}\label{eq03}
y(x)=-\frac{1}{x^2}+\frac{a}{12}-\,\left(\frac{a^2}{240}+\frac{b}{10}\right)\,x^{2}+a_6\,x^4+...
\end{equation}
where $a_6$ is arbitrary constant.

Assuming $a=0$ and $b=0$ we obtain from equation \eqref{eq01} the
equation with leading members that is often called as the reduced
equation

\begin{equation}\label{eq04}
E_1[y]=y_{zz}+6\,y^2=0
\end{equation}

The expansion of solution for equation \eqref{eq04} in the Laurent
series takes the form

\begin{equation}\label{eq05}
y(x)=-\frac{1}{x^2}+a_6\,x^4+...
\end{equation}

Equation \eqref{eq04} has the first integral

\begin{equation}\label{eq06}
I_1[y]=y_{z}^2+4\,y^3-C_1=0
\end{equation}

Substituting \eqref{eq05} into the first integral \eqref{eq06} we
have

\begin{equation}\label{eq07}
I_1[y]=28\,a_{6}+4\,a_{6}^{2}\,x^{6}+4\,a_{6}^{3}\,x^{12}-C_1=0
\end{equation}

From equation \eqref{eq07} follows that if we let $x$ tends to $0$
then $C_1$ goes $28\,a_6$. Therefore arbitrary constant $C_1$ in
the first integral \eqref{eq06} is determined by the arbitrary
constant $a_6$ in the expansion of the general solution of
equation \eqref{eq04} in the Laurent series. However from
expansion \eqref{eq05} we have $a_6\simeq\,y_{xxxx}$. We can
believe that the arbitrary constant $a_6$ corresponds to the pole
of the sixth order because if we use solution $y=-1/x^2$ we obtain
the sixth degree. Taking equation \eqref{eq07} into account one
can note that $C_1$ also corresponds to the sixth order and the
greatest Fuchs index in the expansion of solution in the Laurent
series determines the pole order of members in the first integral
\eqref{eq06}.

\textit{Example 2.} Consider the equation

\begin{equation}\label{eq08}
E_2[y]=y_{zzz}-6\,y^2\,y_{z}=0
\end{equation}

Equation \eqref{eq08} possesses the Painleve property and the
general solution is expressed via the Jacobi elliptic function.
Solution of this equation has the pole of the first order, two
branches and the Fuchs indices $(j_1,\,j_2,\,j_3)=(-1,\,3,\,4)$.
Equation \eqref{eq08} has the first integral

\begin{equation}\label{eq08a}
I_2[y]=y_{zz}-2\,y^3-C_1-0
\end{equation}

Equation \eqref{eq08a} also has the first integral

\begin{equation}\label{eq09}
I_3[y]=y_{z}^2-\,y^4-2\,C_1\,y-C_2=0
\end{equation}

 The general solution of equation \eqref{eq08} can be written in the form of the Laurent series

\begin{equation}\label{eq010}
y(x)=\pm\,\frac{1}{x}+a_3\,x^2+a_4\,x^3+...
\end{equation}
where $a_3$ and $a_4$ are arbitrary constants.

It should be noted that one can say about the pole order of the
arbitrary constants $a_3$ and $a_4$  again because we have
$a_3\simeq\,y_{xx}$ and $a_4\simeq\,y_{xxx}$ from expansion
\eqref{eq010}.

Substituting \eqref{eq010} into the first integral \eqref{eq08} we
obtain

\begin{equation}\begin{gathered}\label{eq010a}
I_2[y]=-4\,{a_3}-6\,{{a_3}}^{2}\,{x}^{3}-12\,{a_3}\,{
a_4}\,{x}^{4}-6 \,{{ a_4}}^{2}\,{x}^{5}-2\,{{
a_3}}^{3}\,{x}^{6}-6\,{{ a_3}}^{2}{ a_4}\,{x}^ {7}-\\
\\
 -6\,{ a_3}\,{{
a_4}}^{2}\,{x}^{8}-2\,{{ a_4}}^{3}\,{x}^{9 }-{C_1}=0
\end{gathered}\end{equation}

Letting $x\rightarrow 0$ in \eqref{eq010a} we obtain $C_1=-4\,a_3
$. We have again from equation \eqref{eq010a} that the arbitrary
constant $C_1$ in the first integral \eqref{eq08a} is determined
by the arbitrary constant $a_3$ that corresponds to the Fuchs
indices. However the arbitrary constant $a_3$ corresponds to the
pole of the third order because $a_3\simeq\,y_{xx}$. We have that
the arbitrary constant $C_1$ also corresponds to the pole of the
third order and we can observe that members of the left hand side
in the first integral \eqref{eq08a} has the pole of the third
order.

Substituting \eqref{eq010} into the first integral \eqref{eq09} at
$C_1=-4\,a_3$ we have

\begin{equation}\begin{gathered}\label{eq011}
I_3[y]=-{ C_2}-10\,{ a_4}+6\,{{ a_3}}^{2}{x}^{2}+8\,{ a_3}{
a_4}\,{x}^{3}+3\,{{ a_4}}^{2}{x}^{4}-4{{ a_3} }^{3}\,{x}^{5}-\\
\\
-12{{ a_3}}^{2}{ a_4}\,{x}^{6}-12{ a_3} \,{{
a_4}}^{2}\,{x}^{7}-\left ({{ a_3}}^{4}+4\,{{ a_4}}^{3}\right
){x}^{8}-4\,{{ a_3}}^{3}{a_4}{x}^{9}-\\
\\
-6\,{{ a_3}}^{2}{{ a_4}}^{2}{x}^{10}-4\,{ a_3}{{
a_4}}^{3}\,{x}^{11}-{{ a_4}}^{4}{x}^{12}=0
\end{gathered}\end{equation}

Letting $x\longrightarrow 0$ in \eqref{eq011} we obtain
$C_2=-10\,a_4$. We can see that the arbitrary constant $C_2$ in
the first integral \eqref{eq09} is determined by the arbitrary
constant $a_4$. The pole order corresponds to the arbitrary
constant $a_4$ equal to four because $a_4\simeq\,y_{xxx}$. We
observe again that the members of the left hand side in the first
integral \eqref{eq09} has the fourth order and corresponds to the
Fuchs index.

These examples allow us to formulate the following theorem.

\begin{theorem}
\label{T:1.1.} Let the nonlinear differential equation in a
polynomial form posses the Painleve property then the pole order
of members in the first integrals of the reduced equation is
determined by the Fuchs indices.
\end{theorem}

\begin{proof} Consider the nonlinear ordinary differential equation in
the polynomial form that possesses the Painleve property and
consequently the reduced equation also has the Painleve property.

We also consider the autonomous ordinary differential equations in
this paper that posses the Painleve property and have the first
integrals in the polynomial form. We know many nonlinear ordinary
differential equations (for example the Painleve equations and the
higher order Painleve equations) that have not got any first
integrals in a polynomial form.

Solution of equation studied can be presented in the Lorent
series. Some coefficients ($n-1$, where $n$ is the order of
equation) of the expansion in the Laurent series are arbitrary
constants. Substituting $y(x)=B_0/x^{p}$ into the leading members
of equation studied we have branches of solutions, that correspond
to different values $(B_0,p)$. As consequence of finding
$(B_0,p,)$ the equation with leading members is satisfied by
solution $y(x)=B_0/x^{p}$.

If equation has the first integral in the polynomial form then the
reduced equation has the first integral in the polynomial form as
well. Substituting $y(x)=B_0/x^{p}$ into the first integral for
the reduced equation we have that is solution of the first
integral when arbitrary constant equals zero.

Substituting the Laurent series for the general solution of
equation with the leading members into the corresponding first
integral we obtain the arbitrary constant in the first integral
that is determined by the arbitrary constants from the Laurent
series. Therefore the arbitrary constant in the first integral of
the nonlinear ordinary differential equation with leading members
is determined by the arbitrary constants in the expansion of
solution in the Laurent series.

However the arbitrary constants in the expansion of the solution
in the Laurent series correspond to the Fuchs indices at
investigation of equation on the Painleve property. Taking the
pole orders of arbitrary constants into consideration (that are
determined by the Fuchs indices) we determine the pole order of
arbitrary constant in the first integral and consequently the pole
order of members in the first integral.
\end{proof}

This theorem allows us to determine the form of the first
integrals of the reduced equation (equation with the leading
members) and to find the first integrals for the more general
equation in the polynomial form.

\section{Algorithm to look for the first integrals
\\of nonlinear differential equations}

Let us use the results of the theorem 2.1 to look for the first
integrals of nonlinear ordinary differential equations. Without
loss of the generality we consider the fourth order ordinary
differential equations. Denote  $y = y_0$, $y_x = y_1$, $y_{xx} =
y_2$, $y_{xxx} = y_3 $, $y_{xxxx} = y_4 $ to simplify
calculations.

Let us assume there is the nonlinear differential equation of the
fourth order in the form

\begin{equation}
\label{eq8} y_4 = E\left( {y_0 ,y_1 ,y_2 ,y_3 } \right)
\end{equation}

Let us suppose there is the first integrals of this equation

\begin{equation}
\label{eq9} P(y_0, y_1, y_2, y_3 ) = K_1
\end{equation}
where $K_1$ and further $K_2$, $K_3$ and $K_4$ are arbitrary
constants.

 Let us obtain the equation for finding of the first integrals of
 nonlinear ordinary differential equations. By definition of the first integral
  of equation \eqref{eq8} we have

\begin{equation}
\label{eq10}\sum_{n=0}^{3}\,y_{n+1}\,\frac{\partial P}{\partial
y_n}\,=\,Q(y_0 ,y_1 ,y_2 ,y_3 ,y_4 )\,\,\left({y_4 - E(y_0 ,y_1
,y_2 ,y_3 )}\right)
\end{equation}
where $Q(y_0,y_1,y_2,y_3,y_4)$ is arbitrary expression depending
on variable $y_0$ and its derivatives.

Taking into account \eqref{eq8} we get from equation \eqref{eq10}

\begin{equation}
\label{Base} \sum_{n=0}^{2}\,y_{n+1}\, \frac{\partial P}{\partial
y_n }\, +\, E(y_0 ,y_1 ,y_2 ,y_3 )\,\frac{\partial P}{\partial y_3
} = 0
\end{equation}

The last equation is the basic equation that can be used to look
for the first integrals of the nonlinear differential equation.
Substituting $E(y_0 ,y_1 ,y_2 ,y_3 )$ from equation \eqref{eq8}
into equation \eqref{Base} we have equation for finding of the
first integrals of equation \eqref{eq8}. To look for the first
integral of equation \eqref{eq8} we have to have additional
information about the form of $P$. This information can be
extracted from the Fuchs indices of equation \eqref{eq8}. Taking
the Fuchs indices into account we can determine the pole order of
the first integral and we can write the polynomial with unknown
coefficients. Substituting the polynomial with unknown
coefficients into equation \eqref{Base} we can find unknown
coefficients solving the algebraic linear equations and obtain the
first integral.

Our approach has the following steps: 1) The Painleve test of
origin equation on the Painleve property and finding the Fuchs
indices; 2) Determination of the reduced equation from the origin
equation. This equation can be found from the origin equation if
we substitute $y_0 = B_{0} / x^p$ into the origin equation; 3)
Writing the polynomial with unknown coefficients taking into
account values of the Fuchs indices; 4) Determination of unknown
coefficients of polynomial and writing the first integral of the
reduced equation; 5) Writing additional polynomials with unknown
coefficients that correspond to the origin equation; 6)
Determination of unknown coefficients of the additional
polynomials and finding the first integral of the origin equation.

\section{The first integrals of equation \eqref{eq1}}

Let us apply our approach to look for the first integrals of the
equation

\begin{equation}
\label{eq111}
 y_{xxxx} - 10y\,y_{xx} - 5y_x^2 + 10y^3 + \beta \left( {y_{xx} - 3y^2}
\right)\, + \,\delta \,y + \mu = 0
\end{equation}

This equation possesses the Painleve property. The general
solution has two branches with $(B_0,p)=(2,2)$, $(B_0,p)=(6,2)$
and the following Fuchs indices $(m_1,\,m_2,\, m_3,\,
m_4)$=$(-1,\, 2,\, 5,\, 8)$ and $(m_1,\, m_2,\, m_3,\,
m_4)$=$(-3,\, -1,\, 8,\,10)$.

The reduced equation can be found from equation \eqref{eq111} at
$\beta=0,\,\delta=0,\,\mu=0$ and takes the form

\begin{equation}
\label{eq20}
 y_{xxxx} - 10y\,y_{xx} - 5y_x^2 + 10y^3 = 0
\end{equation}

Let us find the first integrals of equation \eqref{eq20}. The
greatest Fuchs indices are equal to 8 and 10. One can expect that
the pole order of members for the first integral of equation
\eqref{eq20} is equal to 8 and the pole order of members for
another first integral is equal to 10. Let us write two polynomial
with unknown coefficients that correspond to the poles order 8 and
and 10.

We denote the polynomial $P_{k}^{(j)}$ as polynomial with the k-th
pole order of members and the j-th pole order for solution of the
nonlinear ordinary differential equations.

The first polynomial to look for the first integral of equation
\eqref{eq20} has the form

\begin{equation}
\label{P8} P_{8}^{(2)}= A _0\,y_3\, y_1 + + A_1 y^2_2
  + A_2 y_2 y_0^2 + A_3 y_1^2 y_0   + A_4 y_0^4
\end{equation}
where $A_0,...,A_{4}$ are unknown coefficients that should be
found. One can note that members in the polynomial \eqref{P8} has
the eighth order.

The second polynomial to search the first integral of equation
\eqref{eq20} takes the form

\begin{equation}
\label{P10} P_{10}^{(2)}= A_0 y^2_3  + A _1y_3 y_1y_0 + A_2
y^2_2y_0
 + A_3 y_2 y_1^2    + A_4 y_2 y_0^3   + A_5 y_1^2 y_0^2   + A_6 y_0^5
\end{equation}

Substituting polynomial \eqref{P8} into equation \eqref{Base} and
solving algebraic equations for the unknown coefficients
$A_0,...A_4$ we have the first integral for equation \eqref{eq20}
in the form

\begin{equation}
\label{Ieq1} I_{4}[y]=y_{{x}}y_{{{\it xxx}}}-\frac 12\,{y_{{{\it
xx}}}}^{2}-5\,{y}{y_ {{x}}}^{2}+\frac52\,{y}^{4}-{C_{{1}}}=0
\end{equation}

The expansion of the general solution of equation \eqref{eq20} in
the Laurent series for the first branch takes the form

\begin{equation}
\label{eq20a} y(x)=\frac{2}{{x}^{2}}+{a_2}-\frac32\,{{
a_2}}^{2}{x}^{2}+{ a_5}\,{x}^{3}-\frac52 \,{{
a_2}}^{3}{x}^{4}+\frac34\,{ a_2}\,{ a_5}\,{x}^{5}+{ a_8}\,{x}
^{6}+...
\end{equation}

Substituting solution \eqref{eq20a} into the first integral
\eqref{Ieq1} and letting $x\rightarrow0$ we obtain

\begin{equation}
\label{eq20b}I_{4}[y_{1}]=-567\,a_2^4-648\,a_8-C_1=0
\end{equation}

Equality \eqref{eq20b} confirms our choose of the polynomial
\eqref{P8} and the theorem 2.1.

Later we need the following polynomials

\begin{equation}
\begin{gathered}
\label{P6}P_{0}^{(2)}=A_0,\,\quad\,P_{2}^{(2)}=A_{0}\,y_{0},\,\quad\,P_{3}^{(2)}=A_{0}\,y_{1},\,\quad\,
P_{4}^{(2)}=A_{0}\,y_{2}+A_{1}\,y_{0}^{2},
\\
\\
P_{5}^{(2)}=A_{0}\,y_{3}+A_{1}\,y_{1}y_{0},\,\quad\,
P_{6}^{(2)}=A_{0}\,y_{4}+A_{1}\,y_{2}y_{0}+A_{2}\,y_{1}^{2}+A_{3}\,y_{0}^{3}
\end{gathered}
\end{equation}

Adding to the first integral \eqref{Ieq1} three polynomials with
unknown coefficients in the form $\beta\,P_{4}^{(2)}+\delta\,
P_{3}^{(2)}+\mu\, P_{2}^{(2)} $ we obtain the following first
integral of equation \eqref{eq111}

\begin{equation}
\label{Ieq2} y_{{x}}y_{{{\it xxx}}}-\frac 12\,{y_{{{\it
xx}}}}^{2}-5\,{y}{y_ {{x}}}^{2}+\frac52\,{y}^{4}+\frac
12\,\beta\,\left ({y _{{x}}}^{2}-2{y}^{3}\right )+\frac
12\,\delta\,{y}^{2}+\mu\,y={\it K_{{1}}}
\end{equation}

Substituting polynomial \eqref{P10} into equation \eqref{eq11} and
finding $A_0,...A_6$ we obtain another first integral for the
equation \eqref{eq20} in the form

\begin{equation}
\begin{gathered}
\label{Ieq2a} {y_{{{\it xxxx}}}}^{2}-12\,yy_{{x}}y_{{{\it
xxx}}}-4\,y{y_{{{\it xx}}} }^{2} +2\,{y_{{x}}}^{2}y_{{{\it
xx}}}+20\,{y}^{3}y_{{{\it
xx}}}+\\
\\
+30\,{y} ^{2}{y_{{x}}}^{2}-24\,{y}^{5}-C_{{2}}=0
\end{gathered}
\end{equation}

Adding to \eqref{Ieq2a} $\beta\,P_6+\delta\, P_5+\mu\, P_4 $ and
substituting them into equation \eqref{eq11} again we have the
first integral in the form

\begin{multline}
\label{Ieq2b}
{y_{{{xxxx}}}}^{2}-12\,yy_{{x}}y_{{{xxx}}}-4\,y{y_{{{ xx}}}
}^{2}+2\,{y_{{x}}}^{2}y_{{{ xx}}}+20\,{y}^{3}y_{{{xx}}}+30\,{y}
^{2}{y_{{x}}}^{2}-24\,{y}^{5}+\\
\\
+\beta\,\left(y_{{xx}}-3\,{y }^{2}\right)^2+\delta\,\left(2\,y y
_{{{xx}}}-4\,{y}^{3}-{y_{{x}}}^{2}\right)+2\,\mu\,\left (y_{{{
xx}}}-3\,{y}^{2}\right )=K_{{2}}
\end{multline}

The first integrals \eqref{Ieq2} and \eqref{Ieq2b} can be used to
look for the general solution of equation \eqref{eq111}.

\section{The first integrals of equation \eqref{eq2}}

Let us find the first integrals of the equation

\begin{equation}
\label{eq23}
 y_{xxxx} - 10y\,^2y_{xx} - 10y\,y_x^2 + 6y^5 + \beta \,\left( {y_{xx} -
2y^3} \right)\, + \delta \,y + \mu = 0
\end{equation}

This equation passes the Painleve test. Solution of \eqref{eq23}
has the pole of the first order and four branches with
$(B_0,\,p)=(1,\,1)$, $(B_0,\,p)=(-1,\,1)$, $(B_0,\,p)=(2,\,1)$ and
$(B_0,\,p)=(-2,\,1)$. These branches have two collections of the
Fuchs indices: $(j_1,\,j_2\,j_3,\,j_4)$=$(-1,\,2,\,3,\,6)$ and
$(j_1,\,j_2\,j_3,\,j_4)$=$(-3\,-1,\,6,\,8)$. The greatest Fuchs
indices are equals to six and eight. Therefore we can expect that
members of the first integrals have poles of the sixth and eighth
order.

The reduced equation can be found from equation \eqref{eq23} at
$\beta=0,\,\delta=0,\,\mu=0$ and takes the form

\begin{equation}
\label{eq24}
 y_{xxxx} - 10y\,^2y_{xx} - 10y\,y_x^2 + 6y^5 = 0
\end{equation}

In fact one of the first integral of equation \eqref{eq24} can be
found if we take the pole of the sixth order into account and
multiply equation \eqref{eq24} on $y_{x}$ but we are going to use
our approach and to construct the polynomial with unknown
coefficients that has members with pole of the sixth order. This
polynomial can be written in the form

\begin{equation}\begin{gathered}
\label{eq25}
 P_{6}^{(1)}=a_0\,y_{1}y_{3}+a_1\,y_3y_{0}^{2}+a_2\,y_{2}^{2}+a_3\,{y}_2\,y_{1}y_0+a_4\,y_2y_{0}^{3}
 a_5\,y_{1}^{3}+\\
 \\
 +a_6\,y_{1}^{2}y^{2}_{0}+a_7\,y_{1}y_{0}^{4}+a_{8}\,y_{0}^{6}
\end{gathered}\end{equation}
where $a_0,...,a_{8}$ are unknown coefficients that  should be
found.

Substituting \eqref{eq25} into equation \eqref{Base} and solving
the system of the linear algebraic equations for coefficients
$a_0,...,a_{8}$ we have the first integral of equation
\eqref{eq24} in the form

\begin{equation}
\label{eq26} I_{5}[y]=y_{1}y_{3}-\frac 12\,y_{2}^{2}-5\,y^{2}y_
{1}^{2}+y_{0}^{6}-C_{1}=0
\end{equation}

The expansion of the solution of equation \eqref{eq24} in the
Laurent series of the first branch  takes the form

\begin{equation}\begin{gathered}
\label{eq26a} y(x)=\frac{1}{x}+{a_2}\,x+{a_3}\,{x}^{2}+\frac52\,{{
a_2}}^{2}\,{x}^{3}+\frac53 \,{ a_2}{a_3}\,{x}^{4}+{
a_6}\,{x}^{5}+...
\end{gathered}\end{equation}

Substituting the expansion \eqref{eq26a} into the first integral
\eqref{eq26} and letting $x\rightarrow0$ we have the equality

\begin{equation}
\label{eq26b} I_{5}[y_1]=210\,{{a_2}}^{3}+28\,{{ a_3}}^{2}-84\,{
a_6}-{C_1}=0
\end{equation}

This equality again gives the confirmation of the theorem 2.1.

Later we are also going to use the following polynomials

\begin{equation}\begin{gathered}
\label{Q3} P_{1}^{(1)}=a_{0}y_{0},\,\quad\,
P_{2}^{(1)}=a_{0}y_{1}+a_{1}y_{0}^{2},\,\quad\,
P_{3}^{(1)}=a_{0}y_{2}+a_{1}y_{1}y_{0}+a_{2}\,y_{0}^{3},
\end{gathered}\end{equation}

\begin{equation}\begin{gathered}
\label{Q4}
P_{4}^{(1)}=a_{0}y_{3}+a_{1}y_{2}y_{0}+a_{2}y_{1}^2+a_{3}y_{1}y_{0}^{2}+a_{4}y_{0}^{4}
\end{gathered}\end{equation}

\begin{equation}\begin{gathered}
\label{Q5}
P_{5}^{(1)}=a_{0}y_{4}+a_{1}y_{3}y_{0}+a_{2}y_{2}y_{1}+a_{3}y_{1}^{2}y_{0}+a_{4}y_{1}y_{0}^3+a_{5}y_{0}^5
\end{gathered}\end{equation}

\begin{equation}\begin{gathered}
\label{Q6}
P_{5}^{(1)}=a_{0}y_{4}+a_{1}y_{3}y_{0}+a_{2}y_{2}y_{1}+a_{3}y_{1}^{2}y_{0}+a_{4}y_{1}y_{0}^3+a_{5}y_{0}^5
\end{gathered}\end{equation}

\begin{equation}\begin{gathered}
\label{Q7}P^{(1)}_{7}=a_0\,y_{3}y_{2}+
 a_1\,y_3y_{1}y_0+a_2\,y_{2}^{2}y_{0}
 +a_3\,y_2y_{1}^{2}+a_4\,y_2y_1y_0^{2}+a_5\,y_2y_0^4
 +\\
 \\
 +a_{6}\,y_{1}^{3}y_{0}+a_{7}\,y_{1}^{2}y_{0}^{3}+a_{8}\,{y}_1y_{0}^5+
 a_{9}\,y_{0}^{7}
\end{gathered}\end{equation}

Taking parameters $\beta$, $\delta $ and $\mu$ into account and
adding to \eqref{eq26} the polynomials
$\beta\,P_{4}^{(1)}+\delta\,P_{2}^{(1)}+\mu\,P_{1}^{(1)}$ we have
the first integral of equation \eqref{eq23} in the form

\begin{multline}
\label{eq27}
 y_{{x}}y_{{{\it xxx}}}-\frac 12\,{y_{{{\it xx}}}}^{2}-5\,{y}^{2}{y_
{{x}}}^{2}+{y}^{6}+\frac 12\,\beta\,\left ({y
_{{x}}}^{2}-{y}^{4}\right )+\frac 12\,\delta\,{y}^{2}+\mu\,y={\it
K_{{1}}}
\end{multline}

To find another first integral we use the polynomial with members
of the eighth order pole

\begin{equation}\begin{gathered}
\label{eq28}
 P^{(1)}_{8}=a_0\,y_{3}^{2}+a_1\,y_{3}y_{2}y_{0}+a_3\,{y}_{3}\,y_{1}^2+
 a_4\,y_3y_{1}y_0^{2}+a_5\,y_{2}^{2}y_1+a_6\,y_{2}^{2}y_{0}^{2}+\\
 \\
 +a_7\,y_2y_{1}^{2}y_0+a_8\,y_2y_1y_0^{3}+a_9\,y_2y_0^5
 +a_{10}\,y_{1}^{4}+a_{11}\,y_{1}^{3}y_{0}^2+\\
 \\
 +a_{12}\,y_{1}^{2}y_{0}^{4}+a_{13}\,{y}_1y_{0}^6+
 a_{14}\,y_{0}^{8}
\end{gathered}\end{equation}

Substituting polynomial \eqref{eq28} into equation \eqref{Base}
and taking  equation \eqref{eq24} into account we have another
first integral of equation \eqref{eq24}

\begin{equation}\begin{gathered}
\label{eq29}
y_{3}^{2}-12\,y_{3}y_{1}y_{0}^{2}-4\,y_{2}^{2}y_{0}^{2}+4\,y_{2}y_{1}^{2}y_0
+12\,y_{2}y_{0}
^{5}-y_{1}^{4} +\\
\\
+30\,y_{1}^{2}y^{4}-9\,y_{0}^{8}=C_2
\end{gathered}\end{equation}

Using \eqref{eq29} and adding
$\beta\,P_{6}^{(1)}+\delta\,P^{(1)}_{4}+\mu\,P_{3}^{(1)}$ we have
the first integral of equation \eqref{eq23} in the form

\begin{equation}\begin{gathered}
\label{eq30} {y_{{{xxx}}}}^{2}-12\,{y}^{2}y_{{x}}y_{{{
xxx}}}-4\,{y}^{2}{y_{{{xx}}}}^{2}+4\,y{y_{{x}}}^{2}y_{{{
xx}}}+12\,{y }^{5}y_{{{ xx}}}
-\\
\\
-{y_{{x}}}^{4}+30\,{y}^{4}{y_{{x}}}^{2}-9\,{y}^{8}+\,\beta\left(y_{{xx}}-2y^3\right)^2+\\
\\
+\delta\,\left(2\,yy_{{xx}}-\,
y_{{x}}^2-\,3\,y^4\right)+2\,\mu\left(y_{{xx}}-2y^3\right)=K_{{2}}
\end{gathered}\end{equation}

Equations \eqref{eq111} and \eqref{eq23} can be written as the
Hamiltonian systems and the first integrals found enough to find
the general solutions of these equations.

\section{The first integrals of equation \eqref{eq3}}

Let us find the first integrals of the equation

\begin{equation}
\label{eq31}
 y_{xxxx} + 5\,y_x y_{xx} - 5\,y\,^2y_{xx} - 5\,yy_x^2 + y^5 - \delta =0
\end{equation}

This equation passes the Painleve test. There are four branches of
solutions with $(B_0,\,p)=(1,\,1)$, $(B_0,\,p)=(-3,\,1)$,
$(B_0,\,p)=(-2,\,1)$, $(B_0,\,p)=(4,\,1)$ and three collections of
the Fuchs indices that are $(j_1,\,j_2\,j_3,\,j_4)\,=\,
(-1,\,2,\,3,\,6)$, $(j_1,\,j_2\,j_3,\,j_4)$\,=\,$(-1,\,2,\,6,\,7)$
and $(j_1,\,j_2\,j_3,\,j_4)$\,=\,$(-7,\,-1,\,6,\,12)$. The first
and third branches of solutions have the same Fuchs indices.

Equation with leading members takes the form

\begin{equation}
\label{eq32}
 y_{xxxx}- 5y\,^2y_{xx} + 5y_x y_{xx}- 5y\,y_x^2  + y^5 =0
\end{equation}

There are the Fuchs indices equal six and we can find one of the
first integral of equation \eqref{eq32} using the polynomial
$P_{6}^{(1)}$ from the previous section but we can also obtain
this first integral if we multiply \eqref{eq32} on $y_x$ and
integrating this equality. We have the first integral of equation
\eqref{eq32} in the form

\begin{equation}
\label{eq33} I_{6}[y]=y_3 y_1 - \frac{1}{2}y_2^2 +
\frac{5}{3}y_1^3 - \frac{5}{2}\,y_1^2y_{0}^2 + \frac{1}{6}y_{0}^6
- C_1=0
\end{equation}

By analogy we can find the first integral of equation \eqref{eq31}

\begin{equation}
\label{eq34} y_x y_{xxx} - \frac{1}{2}y_{xx}^2 -
\frac{5}{2}y^2\,y_x^2 + \frac{1}{6}y^6 + \frac{5}{3}y_x^3 - \delta
\,y = K_1
\end{equation}

Substituting the expansions in the Laurent series for the four
branches of the solution into the first integral \eqref{eq33} and
letting $x\rightarrow0$ we have equalities

\begin{equation}
\label{eq34a}I_{6}[y_{1}]={\frac {21}{4}}\,{{ a_2}}^{3}-12\,{{
a_3}}^{2}-84\,{ a_6}-{ C_1}=0
\end{equation}

\begin{equation}
\label{eq34b}I_{6}[y_2]=C_1+168\,{a_6}=0
\end{equation}

\begin{equation}
\label{eq34c}I_{6}[y_3]=-816\,{{ a_2}}^{3}+108\,{{
a_3}}^{2}+168\,{ a_6}-{ C_1}=0
\end{equation}

\begin{equation}
\label{eq34d}I_{6}[y_4]=2184\,a_6-C_1=0
\end{equation}

We have again the confirmation of the theorem 2.1 for the first
integral \eqref{eq33} of equation \eqref{eq32}.

To find another first integral we use the polynomial with members
that have the poles of the twelfth order

\begin{equation}\begin{gathered}
\label{eq35}
 P_{12}^{(1)} = a_0 y_3^3 + \left(a_1 y_2 y_0 + a_2 y_1^2 + a_3 y_1
y_0^2 + a_4 y_0^4 \right)y_3^2 + \\
\\
+\left(a_5 y_0^2+ a_6 y_1\right)y_3 y_2^2 +\left(a_7 y_1^2 y_0 +
a_8 y_1 y_0^3 + a_9 y_0^5 \right)y_3 y_2 +\\
\\
+\left(
a_{10} y_1^4 + a_{11}  y_1^3 y_0^2+ a_{12} y_1^2 y_0^4+ a_{13} y_1 y_0^6 +
a_{14} y_0^8 \right)y_3 + b_0 y_2^4+\\
\\
 + \left(b_1 y_1 y_0 + b_2
 y_0^3\right)y_2^3 + \left(b_3 y_1^3 + b_4 y_1^2 y_0^2 + b_5 y_1 y_0^4 +
 b_6 y_0^6\right)y_2^2 +\\
 \\
 + \left(b_7 y_1^4 y_0 + b_8 y_1^3 y_0^3
 + b_9 y_1^2 y_0^5 + b_{10} y_1 y_0^7 + b_{11} y_0^9\right)y_2 +c_0 y_1^6+ \\
 \\
  + \left(c_1
y_1^4 + c_2 y_1^3 y_0^2
 + c_3 y_1^2 y_0^4 + c_4 y_1 y_0^6 + c_5 y_0^{8}\right)y_1y_0^2+ c_6 y_0^{12}
\end{gathered}\end{equation}
where  $a_0,...,a_{14}$, $b_0 ,...,b_{11}$ and $c_0 ,...,c_6$ are
unknown coefficients that should be found. Substituting this
polynomial into equation \eqref{Base} we have the first integral
of equation \eqref{eq32} in the form

\begin{equation}\begin{gathered}
\label{eq36}
 y_{3}^3 + \left( {\frac{3}{2}y_{0}^4 - 9y_0^2y_1 } \right)\,y_{3}^2 + F_{1}(y_0,y_{1},y_{2})\,y_{3}-
 \frac{15}{8}\,y_{2}^4 + 2y_0^3\,y_{2}^3 + \\
 \\
 +F_{2}(y_0,y_{1})\,y_{2}^2+ F_{3}(y_0,y_{1})\,y_{2}  -
\frac{22}{3}\,y_1^6 +\frac{35}{2}y_0^2\,y_1^5 + \frac{45}{8}y_0^4\,y_1^4-\\
 \\
 - \frac{157}{6}y_0^6\,y_1^3
 + \frac{19}{4}y_0^8 + 3y_0^{10}\,y_1  - \frac{17}{24}y_0^{12} = C_2
\end{gathered}\end{equation}
where

\begin{equation*}\begin{gathered}
F_{1}(y_0,y_{1},y_{2})= \left(\frac{15}{2}y_1 -
3y_0^2\right)\,y_{2}^2+ 3\left(y_1^2 -
2y_0^2y_{1}+y_0^4\right)\,y_0\,y_{2}-\\
\\
 - 7y_1^4- \frac{9}{2}y_0^2y_1^3+
30y_0^4y_1^2-\frac{17}{2}y_0^6\,y_1,
\end{gathered}\end{equation*}

\begin{equation*}
F_{2}(y_0,y_{1})=\frac{25}{2}y_1^3+ 15y_0^4\,y_1 -
\frac{117}{4}y_0^2\,y_1^2- \frac{13}{4}y_0^6,
\end{equation*}

\begin{equation*}
F_{3}(y_0,y_{1})=9y_0\,y_1^4 - 30y_0^3\,y_1^3+ 36y_0^5\,y_1^2  -
18y_0^7y_1 + 3y_0^9
\end{equation*}

Taking into account the first integral \eqref{eq36} and
polynomials $\delta\, P_{7}^{(1)}+\delta^2\,P_{2}^{(1)}$ we have
another first integral in the form

\begin{equation}\begin{gathered}
\label{eq37}
 y_{xxx}^3 + \left( {\frac{3}{2}y^4 - 9y^2y_x } \right)\,y_{xxx}^2 +
 F_{4}(y,y_{x},y_{xx})\,y_{xxx}-
 \frac{15}{8}\,y_{xx}^4 + 2y^3\,y_{xx}^3 + \\
 \\
 +F_{5}(y,y_{x})\,y_{xx}^2+ F_{6}(y,y_{x})\,y_{xx}  -
\frac{22}{3}\,y_x^6 +\frac{35}{2}y^2\,y_x^5 +
\frac{45}{8}y^4\,y_x^4-\\
\\
-\left(\frac{157}{6}y^6- 7\delta \,y\right)\,y_x^3
 + \left(\frac{19}{4}y^8- \frac{27}{2}\delta \,y^3\right)\,y_x^2
 + \\
 \\
 +F_{7}(y)\,y_x  - \frac{17}{24}y^{12}- \frac{9}{2}\delta ^2y^2 =
K_2 \end{gathered}\end{equation}
where

\begin{equation*}\begin{gathered}
F_{4}(y,y_{x},y_{xx})= \left(\frac{15}{2}y_x -
3y^2\right)\,y_{xx}^2+ 3\left(y_x^2 -
2y^2y_{x}+y^4\right)\,y\,y_{xx}-\\
\\
-3\delta\,y_{xx} - 7y_x^4- \frac{9}{2}y^2y_x^3+
30y^4y_x^2-\left(\frac{17}{2}y^6-9\delta y\right)y_x,
\end{gathered}\end{equation*}

\begin{equation*}
F_{5}(y,y_{x})=\frac{25}{2}y_x^3+ 15y^4\,y_x -
\frac{117}{4}y^2\,y_x^2- \frac{13}{4}y^6 - \frac{9}{2}\delta \,y,
\end{equation*}

\begin{equation*}\begin{gathered}
F_{6}(y,y_{x})=9y\,y_x^4 - 30y^3\,y_x^3+ 36y^5\,y_x^2-
9\delta \,y_x^2 + \\
\\
+18\delta \,y^2y_x  - 18y^7y_x
 + 3y^9 - 3\delta y^4,
\end{gathered}\end{equation*}

\begin{equation*}
F_{7}(y)=3y^{10} - 6\,\delta \,y^5+ 3\delta ^2
\end{equation*}

We emphasize that the second branch of solution of equation
\eqref{eq31} has the Fuchs index equal seven but we do not have
this value for the determination of the arbitrary constant $C_1$
in the first integrals found. New first integral can be tolerated
with this Fuchs index that can be found if we use the polynomial
with members of greater pole order .

\section{The first integrals of equations \eqref{eq4} and \eqref{eq5}}

Equations \eqref{eq1}, \eqref{eq2} and \eqref{eq3} have the
polynomial form and we can observe that the Fuchs indices in the
expansion of solutions in the Laurent series determine the first
integrals of nonlinear differential equations. There is question
about other equations with this property. Later we are going to
study the problem of finding of the first integrals in the case
when nonlinear differential equations have the non---polynomial
form. With this aim let us find the first integrals of equation

\begin{equation}\begin{gathered}
\label{eq38} y_{{{\it xxxx}}}-4\,{\frac {y_{{x}}y_{{{\it
xxx}}}}{y}}+ {\frac {{21\,y_{{x}}}^{2}y_{{{\it xx}}}}{{2\,y}^
{2}}}-3\,{\frac {{y_ {{{\it xx}}}}^{2}}{y}}-5\,{\frac
{\delta\,y_{{{ \it xx}}}}{{y}^{2}}}-\,{\frac
{9\,{\,y_{{x}}}^{4}}{2\,{y}^{3}}}+10\,{\frac {\delta\,{y_{{x}}}^{2}}{{y}^{3}}}+\\
\\
+\nu \,{y}^{2}-2\,{\frac {{\delta}^{2}}{{y}^{3}}}+\mu=0
\end{gathered}\end{equation}

Equation \eqref{eq38} passes the Painleve test. Solution has the
first order pole. We get five branches of solutions but two
collections of the Fuchs indices that are
$(j_1,\,j_2\,j_3,\,j_4)$\,=\,$(-1,\,1,\,2,\,4)$ and
$(j_1,\,j_2\,j_3,\,j_4)$\,=\,$(-3,\,-1,\,4,\,6)$.

Consider the reduced equation

\begin{equation}\begin{gathered}
\label{eq39} y_{{{\it xxxx}}}-4\,{\frac {y_{{x}}y_{{{\it
xxx}}}}{y}}+ {\frac {{21\,y_{{x}}}^{2}y_{{{\it xx}}}}{{2\,y}^
{2}}}-3\,{\frac {{y_ {{{\it xx}}}}^{2}}{y}}-\,{\frac
{9\,{\,y_{{x}}}^{4}}{2\,{y}^{3}}}=0
\end{gathered}\end{equation}

Using values for the Fuchs indices we can find the form of the
first integrals. We take the expression in the form
$P_{8}^{(1)}/y_{0}^{4}$ to obtain the value that corresponds to
the order four of the Fuchs indices. We found the first integral
of equation \eqref{eq39} in the form

\begin{equation}
\label{eq40} {\frac {y_{{x}}y_{{{\it xxx}}}}{{y}^{2}}}-\frac
12\,{\frac {{y_{{{\it xx}}}}^{2}}{{y}^{2}}}-2\,{\frac
{{y_{{x}}}^{2}y_{{ { xx}}}}{{y}^{3}}}+{\frac {9}{8}}\,{\frac
{{y_{{x}}}^{4}}{{y}^{4}}} ={C_{{1}}}
\end{equation}

Using additional polynomials
$(\nu\,P_{5}^{(1)}+\delta\,P_{4}^{(1)}+\mu\,P_{3}^{(1)}+\delta^{2}P_{0}^{(1)})/y_{0}^{4}$
we obtain the first integral of the equation \eqref{eq38}

\begin{equation}
\label{eq41} {\frac {y_{{x}}y_{{{\it xxx}}}}{{y}^{2}}}-\frac
12\,{\frac {{y_{{{\it xx}}}}^{2}}{{y}^{2}}}-2\,{\frac
{{y_{{x}}}^{2}y_{{ {\it xx}}}}{{y}^{3}}}+{\frac {9}{8}}\,{\frac
{{y_{{x}}}^{4}}{{y}^{4}}} -\frac 52\,{ \frac
{\delta\,{y_{{x}}}^{2}}{{y}^{4}}}+\frac 12\,{\frac
{{\delta}^{2}}{{y} ^{4}}}-{\frac {\mu}{y}}+\nu\,y={\it K_{{1}}}
\end{equation}

Taking the value of the Fuchs indices equal six into account and
using the expression with unknown coefficients in the form
$P_{12}^{(1)}/y_{0}^{6}$ we have the first integral of the
equation \eqref{eq39} in the form

\begin{equation} \label{eq42} {\frac {{y_{{{\it
xxx}}}}^{2}}{{y}^{2}}}-6\,{\frac {y_{{x}}y_{ {{\it xx}}}y_{{{\it
xxx}}}}{{y}^{3}}}+3\,{\frac {{y_{{x}}}^{3}y_{{{\it xxx}}}}
{{y}^{4}}}+9\,{\frac {{{y_{{x}}}^{2}y_{{{\it
xx}}}}^{2}}{{y}^{4}}}-9\, {\frac {{y_{{x}}}^{4}y_{{{\it
xx}}}}{{y}^{5}}}+\frac 94\,{\frac {{y_{{x}}}^ {6}}{{y}^{6}}}=C_2
\end{equation}

Using equation \eqref{eq42} and the additional polynomials
$(\nu\,P_{9}^{(1)}+\delta\,P_{8}^{(1)}+\mu\,P_{7}^{(1)}+
\delta\,\nu\,P_{5}^{(1)}+\delta^{2}\,P_{4}^{(1)}+\delta\,\mu\,P_{3}^{(1)}+\delta^{3}P_{0}^{(1)})/y_{0}^{6}$
we get another first integral of equation \eqref{eq38} in the form

\begin{multline}
\label{eq43} {\frac {{y_{{{\it xxx}}}}^{2}}{{y}^{2}}}-6\,{\frac
{y_{{x}}y_{ {{\it xx}}}y_{{{\it xxx}}}}{{y}^{3}}}+3\,{\frac
{{y_{{x}}}^{3}y_{{{\it xxx}}}} {{y}^{4}}}+9\,{\frac
{{{y_{{x}}}^{2}y_{{{\it xx}}}}^{2}}{{y}^{4}}}-9\, {\frac
{{y_{{x}}}^{4}y_{{{\it xx}}}}{{y}^{5}}}+\frac 94\,{\frac
{{y_{{x}}}^
{6}}{{y}^{6}}}-\\
\\-2\,\delta\,\left (3\,{\frac {y_{{x}}y_{{{\it xxx}}}}{{y}^{4}}}+
{\frac {{y_{{{\it xx}}}}^{2}}{{y}^{4}}}
-10\,{\frac {{y_{{x}}}^{2}y_{{{\it xx} }}}{{y}^{5}}}+{\frac
{19}{4}}\,{\frac {{y_{{x}}}^{4}}{{y} ^{6}}}\right )-2\,{\frac
{{\delta}^{3}} {{y}^{6}}}+2\,{\frac
{\delta\,\mu}{{y}^{3}}}+6\,{\frac
{\delta\,\nu}{y }}+\\
\\
+2\,\nu\,y_{{{\it xx}}}-3\,{\frac {\nu\,{y_{{x}}}^{2}}{y
}}+2\,{\frac {\mu\,y_{{{\it xx}}}}{{y}^{2}}}-{\frac
{\mu\,{y_{{x}}}^{2 }}{{y}^{3}}}-{\delta}^{2}\left (4\,{\frac
{y_{{{\it xx}}}}{{y}^{5}}}-11\,{\frac
{{y_{{x}}}^{2}}{{y}^{6}}}\right )={\it K_{{2}}}
\end{multline}

Now let us find the first integrals of the equation

\begin{equation}\begin{gathered}
\label{eq44}
 y_{{{\it xxxx}}}-2\,{\frac {y_{{{
xxx}}}y_{{x}}}{y}}-\,{\frac {3\,{y_{{{ xx}}}}^{2}}
{2\,y}}+2\,{\frac {y_{ {{
xx}}}{y_{{x}}}^{2}}{{y}^{2}}}-5\,{y}^{2}y_{{{\it xx}}}
-\frac52\,y{y_{{x}}}^{2} +\frac52\,{y}^{5}-\\
\\
-{\beta}{y }^{3}+\mu\,y=0
\end{gathered}\end{equation}

Equation passes the Painleve test and has four branches of
solution. These branches have two collections of the Fuchs indices
that are $(j_1,\,j_2\,j_3,\,j_4)$\,=\,$(-1,\,1,\,3,\,5)$ and
$(j_1,\,j_2\,j_3,\,j_4)$\,=\,$(-2,\,-1,\,5,\,6)$. We can expect
that the first integrals of the reduced equation

\begin{equation}\begin{gathered}
\label{eq45}
 y_{{{\it xxxx}}}-2\,{\frac {y_{{{
xxx}}}y_{{x}}}{y}}-\,{\frac {3\,{y_{{{ xx}}}}^{2}}
{2\,y}}+2\,{\frac {y_{ {{
xx}}}{y_{{x}}}^{2}}{{y}^{2}}}-5\,{y}^{2}y_{{{\it
xx}}}-\frac52\,y{y_{{x}}}^{2} +\frac52\,{y}^{5}=0
\end{gathered}\end{equation}
can be obtained in the form of the expressions that correspond to
the Fuchs indices equal five and six. Taking
$P_{7}^{(1)}/y_{0}^{2}$ into account we obtain the first integral
of equation \eqref{eq45} in the form

\begin{equation}
\label{eq46} {\frac {y_{{x}}y_{{{\it xxx}}}}{y}}-\frac12\,{\frac
{{y_{{{xx}}}}^{2}}{y}}-{\frac {{y_{{x}}}^{2}y_{{{ xx}}
}}{{y}^{2}}}-\frac 52\,y{y_{{x}}}^{2 }+\frac 12\,{y}^{5}={
C_{{1}}}
\end{equation}

Using the first integral \eqref{eq46} and the additional
polynomials $(\beta\,P_{5}^{(1)}+\mu\,P_{3}^{(1)})/y_{0}^{2}$ we
get the first integral of equation \eqref{eq44} in the form

\begin{equation}
\label{eq47} {\frac {y_{{x}}y_{{{\it xxx}}}}{y}}-\frac12\,{\frac
{{y_{{{\it xx}}}}^{2}}{y}}-{\frac {{y_{{x}}}^{2}y_{{{\it xx}}
}}{{y}^{2}}}-\frac 52\,y{y_{{x}}}^{2 }+\frac 12\,{y}^{5}-\frac
13\,{\beta}{y}^{3}+\mu\,y={\it K_{{1}}}
\end{equation}

To find another first integral we take the expression
$P_{10}^{(1)}/y_{0}^{4}$ into consideration. We obtain

\begin{equation}\begin{gathered}\label{eq48}
{\frac {{y_{{{\it xxx}}}}^{2}}{{y}^{2}}}-2\,{\frac {y_{{x}}y_{
{{\it xx}}}y_{{{\it xxx}}}}{{y}^{3}}}-8\,y_{{{\it
xxx}}}y_{{x}}-\frac 13\,{\frac {{y_{{{
xx}}}}^{3}}{{y}^{3}}}+{\frac {{y_{{x}}}^{2 }{y_{{{
xx}}}}^{2}}{{y}^{4}}}+11\,{\frac {{y_{{x}}}^{2}y_{{{
xx}}}}{y}}-\\
\\
-{y_{{{xx}}}}^{2}+5\,{y}^{3}y_{{{
xx}}}+10\,{y}^{2}{y_{{x}}}^{2}-\frac{10}{3}\,{y} ^{6}={C_{{2}}}
\end{gathered}\end{equation}

Taking  the first integral \eqref{eq48} and polynomials
$(\mu\,P_{6}^{(1)}+\beta\,P_{4}^{(1)})/y_{0}^4$ into account we
get another first integral of equation \eqref{eq44} in the form

\begin{equation}\begin{gathered}\label{eq49}
{\frac {{y_{{{\it xxx}}}}^{2}}{{y}^{2}}}-2\,{\frac {y_{{x}}y_{
{{\it xx}}}y_{{{\it xxx}}}}{{y}^{3}}}-\frac 13\,{\frac {{y_{{{
xx}}}}^{3}}{{y}^{3}}}+{\frac {{y_{{x}}}^{2 }{y_{{{
xx}}}}^{2}}{{y}^{4}}}+11\,{\frac {{y_{{x}}}^{2}y_{{{
xx}}}}{y}}-\\
\\
-8\,y_{{{\it xxx}}}y_{{x}}-{y_{{{xx}}}}^{2}+5\,{y}^{3}y_{{{
xx}}}+10\,{y}^{2}{y_{{x}}}^{2}-\frac{10}{3}\,{y} ^{6}+2\,{\frac
{\mu\,y_{{{
xx}}}}{y}}-4\,\mu\,{y}^{2}-\\
\\
-2\,{\beta}yy_{{{\it
xx}}}+2\,{\beta}{y_{{x}}}^{2}+2\,{\beta}{y}^{4}={ K_{{2}}}
\end{gathered}\end{equation}

We have obtained the first integrals for several nonlinear
ordinary differential equations. All these first integrals have
members with the pole orders that are determined via the Fuchs
indices.

\section{General solution of equation \eqref{eq1}}

Let us find the general solution of equation \eqref{eq1} using the
first integrals \eqref{Ieq2} and \eqref{Ieq2b}. At $\beta =0$
equation was studied before \cite{16,17,18}. Here let us present
the general solution in the case $\beta\neq\,0$. Denote new
variables

\begin{equation}\label{eq50}
M=y_{{xx}}-3\,y^2-\frac12\,\delta,
\end{equation}

\begin{equation}\label{eq51}
N=y\,y_{{xx}}-\frac12\,y_{{x}}^2-3\,y^3-\frac12\,\mu
\end{equation}
then we can write the first integrals \eqref{Ieq2} and
\eqref{Ieq2b} in the form

\begin{equation}
\label{eq52} y_{{x}}M_{{x}}-\left(\, y^2+\frac 12\,\delta-\beta\,y
\right)\, M\,-\,\left(\, \beta+2y \right)\,N-\frac 12\,M^2+\frac
12\,\beta\,\delta\,y=K_{{1}}
\end{equation}

\begin{equation}
\label{eq53}
M_{{x}}^2+\beta\,M^2-4\,M\,N+\beta\,\delta\,M=K_{{2}},
\end{equation}

Let $P(t)$ be the curve of the genus two in the form

\begin{equation*}
P(t) = t^5 + m_0\, t^4 + m_1\, t^3 + m_2\, t^2 + m_3\, t + m_4
\end{equation*}
where $m_{0},...,m_{4}$ are unknown coefficients.

Let us assume that $y$ have the form

\begin{equation}\label{eq54}
y = \frac{1}{2}(u(x) + v(x) - \beta ),\qquad M(x) =
\frac{1}{2}u(x)\,v(x)
\end{equation}
where $u(x)$ and $v(x)$ satisfy the following equations

\begin{equation}
\label{eq55} (u - v)\,u_x \, = \sqrt {P(u)}, \qquad(u - v)\,v_x \,
= - \sqrt {P(v)}
\end{equation}

Substituting \eqref{eq54} and \eqref{eq55} into the first
integrals \eqref{eq52} and \eqref{eq53} we obtain $P(t)$ in the
form

\begin{equation}
\label{eq56}
 P(t) = t^5 - 3\beta \,t^4 + (3\beta ^2 + 2\delta )\,t^2 - 4\mu \,t^2 +  2\,(\beta ^2\delta
+ 4K_1 )\,t + 4K_2
\end{equation}

The system of equations \eqref{eq55} can be written in the form

\begin{equation}\begin{gathered}
\label{eq57} I_{0}(u(x))+I_{0}(v(x))=
K_3,\,\,\,\quad\,\,\,I_{1}(u(x))+I_{1}(v(x))= x+K_4
\end{gathered}\end{equation}
where

\begin{equation}\begin{gathered}
I_{0}(u(x))=\int\limits_\infty ^{u(x)} {\frac{dt}{\sqrt {P(t)} }},
\,\,\,\quad\,\,\, I_{1}(u(x))=\int\limits_\infty ^{u(x)}
{\frac{t\,dt}{\sqrt {P(t)}}}, \end{gathered}\end{equation}

The general solution of equation \eqref{eq1} is expressed via the
hyperelliptic functions taking equations \eqref{eq56} and
\eqref{eq57} into account.

\section {Conclusion}

In this paper we considered the problem of finding of the first
integral for nonlinear ordinary differential equations. We have
observed that arbitrary constant of the first integral is
determined by the arbitrary constants in the expansion of the
general solution in the Laurent series. However the arbitrary
constants for the general solution in the Laurent series have the
pole order that correspond to the Fuchs indices. We have found the
link between the pole order of the members in the first integrals
and the Fuchs indices of the expansion in the Laurent series. This
observation allowed us to present the algorithm to look for the
first integrals of nonlinear differential equations that posses
the Painleve property. Algorithm was applied to look for the first
integrals of five nonlinear ordinary differential equations.

\section {Acknowledgments}

This work was supported by the International Science and
Technology Center under Project No 1379-2.

\end{document}